\documentclass[submission,copyright,creativecommons]{eptcs}
\providecommand{\event}{ICLP/LPNMR-DC 2024} 
\usepackage{amsmath}
\usepackage{amssymb}
\usepackage{fancyvrb}
\usepackage{listings}
\usepackage{wrapfig}
\usepackage{mathrsfs}
\usepackage{enumitem}

\usepackage{iftex}
\usepackage{graphicx}
\usepackage{setspace}

\newcommand{\asp}[1]{\mbox{$\mathtt{#1}$}}

\newcommand{\naf}{{\mathtt{not}}\,}
\ifpdf
  \usepackage{underscore}         
  \usepackage[T1]{fontenc}        
\else
  \usepackage{breakurl}           
\fi

\title{Bridging Deep Learning and Logic Programming for Explainability through ILP}
\author{Talissa Dreossi
\institute{University of Udine\\ Udine, Italy}
\email{talissa.dreossi@uniud.it}
}
\def\titlerunning{Bridging Logic Programming and 
Deep Learning for Explainability through ILASP}
\def\authorrunning{T. Dreossi}
\begin{document}
\maketitle
\vspace{-4mm}
\begin{abstract}
My research explores integrating deep learning and logic programming to set the basis for a new
generation of AI systems. By combining neural networks with Inductive Logic Programming (ILP), the goal is to construct systems that make accurate predictions and generate comprehensible rules to validate these predictions. Deep learning models process and analyze complex data, while ILP techniques derive logical rules to prove the network's conclusions. Explainable AI methods, like eXplainable Answer Set Programming (XASP), elucidate the reasoning behind these rules and decisions. The focus is on applying ILP frameworks, specifically ILASP and FastLAS, to enhance explainability in various domains. My test cases span weather prediction, the legal field, and image recognition. In weather forecasting, the system will predict events and provides explanations using FastLAS, with plans to integrate recurrent neural networks in the future. In the legal domain, the research focuses on interpreting vague decisions and assisting legal professionals by encoding Italian legal articles and learning reasoning patterns from Court of Cassation decisions using ILASP. For biological laboratories, we will collaborate with a research group to automate spermatozoa morphology classification for Bull Breeding Soundness Evaluation using YOLO networks and ILP to explain classification outcomes. This hybrid approach aims to bridge the gap between the high performance of deep learning models and the transparency of symbolic reasoning, advancing AI by providing interpretable and trustworthy applications.
\end{abstract}

\section{Introduction}

The integration of deep learning and logic programming could be a promising approach to creating more interpretable and robust artificial intelligence systems. My research aims to explore and develop a hybrid framework that leverages the strengths of both paradigms. By combining neural networks with ILP \cite{DBLP:journals/ngc/Muggleton91,cropper2022inductive}, the goal is to construct systems capable of not only making accurate predictions but also generating comprehensible rules that validate these predictions. The proposed framework begins with the application of deep learning models, such as neural networks, to process and analyze complex data. Once the neural network produces an output, ILP techniques will be employed to derive logical rules that substantiate the network's conclusions. We also try to apply ILP directly to the data even if often they are too complex or too wide and this limits the capabilities of ILP systems to learn accurate rules. To further enhance the interpretability and reliability of the system, explainable AI methods in logic programming \cite{FANDINNOSCHULZ2019}, such as eXplainable Answer Set Programming (XASP) \cite{alviano2023advancements}, are utilized to elucidate the reasoning behind the derived rules and the decision-making process. I am testing this idea on different fields: weather forecasting, legal judgements and image recognition.

In weather forecasting, accurate predictions are vital for mitigating severe weather impacts, protecting lives, and supporting agriculture, transportation, and disaster management. While traditional methods of weather prediction have greatly benefited from neural networks and deep learning techniques, which offer impressive accuracy, these methods often operate as black boxes, making their predictions hard to understand. This lack of explainability can undermine user trust, impede validation by domain experts, and hinder model refinement. To address these challenges, the research I am pursuing aims to develop a reliable system capable of predicting weather events while also providing explanations. The idea, that was accepted in LPNMR \cite{lpnmrpaper}, is to use FastLAS \cite{fastlas} which is able to generate an optimal subset of possible solutions. FastLAS can quickly identify the best-fitting hypotheses, so that we are then able to translate the learned rules in a human readable way, as a meteorologist would do. 

Similarly, in the legal domain, AI can address complex issues such as the interpretation of vague legal decisions. In facts, legal argumentation often involves vagueness which originates from semantic indeterminacy even when information is available. This branch of my research, which had a recent publication  \cite{cilc2024}, focuses on that while also trying to construct a tool able to assist judges and lawyers in giving an explanation of the final judgement. In particular, I am using some articles of Italian law to test the outcome. In fact, for instance, according to Italian law, street theft ("furto con strappo") can be classified as either an aggravated form of theft or as robbery ("rapina"), depending on factors such as the presence of violence. To tackle this ambiguity problem, we employ Answer Set Programming (ASP) and ILASP \cite{ilasp}. The relevant articles of the Italian legal code are encoded as a set of ASP rules, while decisions from the Court of Cassation are used to make ILASP learn judges' reasoning patterns.

Finally, AI can be employed to speed up human activities in biological laboratory. Indeed, morphological characteristics of bull spermatozoa are typically assessed visually using bright field microscopy following eosin-nigrosine staining, for the Bull Breeding Soundness Evaluation (BBSE). However, this process is time-consuming and demands experienced personnel to achieve reliable results. Given the increasing adoption of genomic selection schemes for young bulls, whose semen is destined for the artificial insemination industry, there is a growing need for a more standardized technique to analyze semen quality. This need is particularly pressing for evaluating spermatozoa abnormalities that impact semen freezing suitability and fertilizing capacity, which are critical due to the widespread use of frozen-thawed semen. Therefore, I am currently developing an AI system for the automated classification of microscope-acquired images of spermatozoa (the study is at the beginning but we are going to present a poster at the European Federation of Animal Science (EEAP) in September). We will employ neural networks, specifically YOLO networks, which can learn and extract relevant features from complex visual data to perform object detection on spermatozoa. This approach will enable us to classify spermatozoa morphology, identifying normal spermatozoa as well as primary and secondary abnormalities. After this initial phase, the plan is to integrate ILP to learn how to identify different morphological characteristics, thereby providing explanations for why specific spermatozoa are classified as abnormal or not. 

The contribution is organized as follows: in Section 2 we briefly introduce the concepts of ILP, focusing on ILASP and FastLAS, and of neural networks, deeply on CNNs, RNN and YOLO networks; an overview of the existing literature is reported in Section 3; Section 4 shows the goal of my research and is followed by Section 5 with an view on the current status and results accomplished of the research.
\vspace{-20pt}

\section{Background}
In this section I am going to explain the main concept that concern my research interests.
\vspace{-3mm}
\subsection{Answer Set Programming} 

Answer Set Programming (ASP) is a declarative programming paradigm born for non-monotonic reasoning and, thanks to the efficiency of ASP solvers,  widely used for modeling and solving difficult combinatorial problems. 
An ASP program is composed by: 
\begin{itemize}[itemsep=-0.5mm]
    \item 
    atoms, generally writings of the form \( p(a_1, \ldots, a_n) \), where \( p \) is a predicate symbol and \( a_i \) are constant or variable symbols with $i \in [1,n]$. Intuitively, such an atom states that the elements represented by \( a_1, \ldots, a_n \) enjoy the property denoted by \( p \);
\item
    rules $r$ of the form $H \leftarrow A_1, \ldots , A_n, \naf B_1, \ldots, \naf B_m$, where $H$, $A_i$, and $B_j$ are atoms of the form $p(t_1,\ldots,t_n)$ where $p$ is a predicate symbol, $t_i$ are constant or variable symbols, and $n\geq 0$.
\end{itemize}

A literal is either an atom $A$ or its default negation $\naf A$,  i.e., a \emph{naf-literal}.
Rules with empty body are called facts, while rules without a head (corresponding to the case $H={\tt false}$) are called constraints or denials. If a rule $r$ has not variable symbols, it is said to be \emph{ground}.  A set of ground atoms $S$ is a stable model of a program $P$ if it is the unique minimum model of the reduct $P^S$ of $P$. The reduct $P^S$ is obtained from $P$ by removing all rules whose body is not satisfied by the atoms in $S$, and removing all naf-literals from the remaining rules~\cite{gelfond1988stable}. An ASP solver can determine whether a program $P$ has any stable models and, if so, compute the set $AS(P)$ of all such models. A key property of stable models is that if an atom $A$ belongs to a stable model $S$, there must be at least one rule $r$ in the ground version of $P$ whose body is satisfied by $S$ and whose head is $A$. This rule $r$ provides an explanation for the truth of $A$, meaning that we not only know $A$ is true, but also the reason for its truth even if multiple alternative explanations exist.

The ASP language supports various syntactic extensions. One notable extension introduces controlled forms of non-determinism in programs, such as \emph{choice rules} and \emph{cardinality constraints}. Additionally, \emph{built-in atoms} of the form ~$E_1 \mathbin{op} E_2$~ (where $E_1$ and $E_2$ are expressions with numerical constants, variables, and arithmetic operators, while $\mathbin{op} \in \{<, \leq, =, \neq, >, \geq\}$) can be used in rule bodies to model arithmetic comparisons between numerical expressions. During the grounding phase, the variables in these expressions are replaced by constants, and the expressions are evaluated and replaced by {\tt true/false}.
\vspace{-4mm}

\subsection{Inductive Logic Programming}

Inductive Logic Programming (ILP) \cite{DBLP:journals/ngc/Muggleton91} is a subfield of machine learning that focuses on learning logical rules from examples and background knowledge. The objective is to discover a hypothesis (a set of logical rules) that explains the given examples within the context of the provided background knowledge (a set of rules). Various ILP frameworks have been proposed in the literature~\cite{law2018inductive}. My research is focused on \emph{Learning from Answer Sets} (LAS), a state-of-the-art ILP framework designed for learning from noisy examples~\cite{law2018inductive}.

LAS \cite{ilasp} is a paradigm for learning answer set programs. It is known that in ASP there can be zero or many answer sets of a program. For this reason we can talk about \emph{brave entailment} ($\models_b$) and \emph{cautious entailment} ($\models_a$): an atom $a$ is bravely entailed by a program $P$ if and only if at least one answer set $P$ contains $a$, while it is cautiously entailed if every answer set contains it. A learning problem in LAS is called a LAS \emph{task}. Specifically, a LAS task is a tuple $T= \langle B, S_M, E\rangle$, where $B$ is an ASP program known as the \emph{background knowledge}, $S_M$ is a set of rules that form the \emph{hypothesis space}, and $E$ is the set of \emph{examples}. 
To avoid the explicit introduction of a (huge) hypothesis space, it is defined through a \emph{mode bias}, which consists of a pair of sets of mode declarations $\langle M_{h}, M_{b} \rangle$. Here, $M_{h}$ (head mode declarations) specify which predicates can appear in the head of a rule, and $M_{b}$ (body mode declarations) specify which predicates can appear in the body of a rule. A mode declaration is a literal whose arguments are either $\asp{var(t)}$ or $\asp{const(t)}$, where $\asp{t}$ is a type. Informally, a literal is \emph{compatible} with a mode declaration $m$ if it can be constructed by replacing every instance of $\asp{var(t)}$ in $m$ with a variable of type $\asp{t}$ and every $\asp{const(t)}$ in $m$ with a constant of type $\asp{t}$.

Examples in LAS are based on the notion of a \emph{partial interpretation}. A partial interpretation $e_{pi}$ is a pair of sets of ground atoms $\langle e^{inc}, e^{exc}\rangle$, where $e^{inc}$ and $e^{exc}$ are referred to as the \emph{inclusions} and \emph{exclusions} sets respectively. An interpretation $I$ is said to \emph{extend} a partial interpretation $e_{pi}$ if and only if $e^{inc} \subseteq I$ and $e^{exc}\cap
I = \emptyset$.  Unlike conventional ILP, the LAS framework allows for context-dependent examples. 
In real world setting, data are \emph{noisy}. Noise can be captured by allowing examples to be weighted with a notion of \emph{penalty}. A \emph{weighted context-dependent partial interpretation} (WCDPI) $e$ is a tuple  $\langle e_{id}, e_{pen}, e_{cdpi}\rangle$, where $e_{id}$ is an identifier for $e$, $e_{pen}$ is a positive integer, called a \emph{penalty}, $e_{cdpi}$ is a context-dependent partial interpretation. A task is called \emph{noisy LAS} task when examples are WCDPIs. Formally, a \emph{Noisy LAS} task is a tuple $T^{noise}=\langle B, M, E\rangle$ where $B$ is an ASP program, $S_M$ is the search space, and $E$ is a finite set of WCDPIs. A hypothesis $H \subseteq S_M$ is an inductive solution of $T^{noise}$ iff $\forall e \in E$, $B\cup H$ accepts $e$. We denote with $\mathcal{T}^{noise}$ the set of all Noisy LAS tasks.
If a hypothesis does not accept an example, it \emph{pays the penalty}, which contributes to the overall \emph{cost} of the hypothesis. A scoring function \(\mathcal{S}\) assigns a positive real number as score to any ASP program and noisy task \(T^{noise}\). The goal of a \emph{noisy} LAS tasks is to find an \emph{optimal} hypothesis that minimizes the cost over a given hypothesis space and WCDPI examples. Precisely, a hypothesis \(H\) is considered the best answer to a noisy LAS task \(T^{noise}\) according to \(\mathcal{S}\) if \(H\) solves \(T^{noise}\) and no other solution \(H'\) has a better score than \(H\).

FastLAS~\cite{fastlas} is a noisy LAS system. It supports user-defined scoring functions, allowing domain-specific optimisation criteria to be used to bias the search; for example, scoring functions can be used to bias towards the cheapest, least risky, safest or most secure set of rules. Furthermore, continuous data types (such as real numbers) are common in machine learning, but many ILP approaches are unable to deal with such data types (without discretisation). FastLAS supports numeric data types, together with binary comparisons over these types.

On the other hand, ILASP is also a framework designed to learn answer set programs from examples and background knowledge. One of its key features is the ability to handle both noise-free and noisy data, making it robust in real-world applications. It infer new rules forcing positive examples to be extended by at least one answer set, while negative examples have not to be extended by any answer set. ILASP uses a \emph{Conflict-Driven ILP} (CDILP) process \cite{law2023conflict} and, unlike FastLAS, it is also able to learn recursive hypotheses, enabling the modeling of complex relationships within the data.
\vspace{-4mm}

\subsection{Artificial Neural Networks}
Regarding Artificial Neural Networks (ANNs), there are numerous articles and books in the literature from which to retrieve information. Therefore, we will cite just the work of Lecun et al. \cite{lecun2015deep} and the survey of Dong et al. \cite{dong2021survey}. ANN is a computational model inspired by biological neural networks. It is composed by interconnected neurons, functioning as processing units, and connections with adjustable weights. Through training algorithms, these weights are tuned based on input data, ending in a discriminating function capable of classifying various inputs. The network's structure, as its biological counterpart, evolves during training and recall phases to optimize classification performance. Every network is constructed by layers of neurons. While the basic concept of ANNs involves interconnected layers of neurons that process and learn from data, various specialized architectures have been developed to address specific types of tasks and data structures more effectively. In this section, we will give a brief description of Convolutional Neural Network (CNN), Recurrent Neural Network (RNN), and finally You Only Look Once Network (YOLO).
\vspace{-4mm}
\paragraph{CNN}
Convolutional Neural Networks (CNNs) are primarily designed for processing structured grid data, such as images. They are particularly effective for image recognition, classification, and segmentation tasks. The convolutional layer is the core building block of a CNN. It applies a set of filters (or \emph{kernels}) to the input image, which slide across the width and height of the input. This operation produces a feature map that highlights the presence of certain features in the input, such as edges, textures, or patterns. During this process, the filter performs element-wise multiplication with the pixel values within the patch.
\vspace{-4mm}
\paragraph{Recurrent Neural Network}
Recurrent Neural Networks (RNNs) are particularly well-suited for sequence modeling tasks. Unlike traditional feedforward neural networks, which process inputs independently, RNNs have an internal memory that allows them to capture temporal dependencies in sequential data. Their basic architecture  consists of three main components: input (at each time step $t$, the RNN layer receives an input vector $x^{(t)}$, representing the input at that time step); hidden State (the RNN layer maintains a hidden state vector $h^{(t)}$ that encapsulates the network's memory of past inputs up to time step $t$); recurrent connection (at each time step, the RNN computes the hidden state $h^{(t)}$ based on the current input $x^{(t)}$ and the previous hidden state $h^{(t-1)}$); output (the RNN layer may produce an output $y^{(t)}$ at each time step).

\vspace{-4mm}
\paragraph{YOLO}
You Only Look Once (YOLO) is a state-of-the-art, real-time object detection algorithm introduced in 2015 by Redmon et al. \cite{redmon2016you}. Unlike traditional object detection methods that use a sliding window approach or region proposal networks, YOLO treats object detection as a single regression problem, directly predicting class probabilities and bounding box coordinates from the entire image in one evaluation.  Mathematically, YOLO divides an input image into an $S \times S$ grid. Each grid cell is responsible for predicting $B$ bounding boxes and their corresponding confidence scores. Each bounding box prediction consists of five components: $(x, y, w, h, confidence)$. $x, y, w$ and $h$ correspond to the coordinates and dimensions of the box, while confidence score \cite{jiang2022review} indicates how likely the box contains an object and how accurate the box is.

\section{Related Works}
In this section, it is explored the state-of-art and the related works in each field relevant to my research.

Concerning weather prediction, traditional methods such as numerical weather prediction models \cite{Kalnay2002}, supplemented by statistical and machine learning models \cite{rasp2018deep,mcgovern2017using,weyn2019can,weyn2020improving,li2024generative}, achieve high accuracy. However, understanding models like neural networks remains challenging. Explainable AI (XAI) techniques are being developed to address this issue. For instance, Labe et al. \cite{labe2023changes} used XAI with ANNs to explain rising summer temperatures in the USA. The idea of my research is, instead, to use ILP since it is able to give explainable results. I am specifically interested in ILASP, which have been already exploited in many study \cite{law2014inductive} and FastLAS \cite{baugh2023neuro,cunnington2023symbolic}. FastLAS, effective in prioritizing specific rules over general ones \cite{policyfastlas}, has also been integrated with neural networks \cite{cunnington2023ffnsl}. Within methods providing explainability, Alviano et al. \cite{alviano2023advancements} and Cabalar et al. \cite{cabalar2014causal} contributed with some methods such as XASP.

In the legal field, Allen \cite{Allen57} pioneered legal document interpretation using symbolic logic. Then, Kowalski and Sergot \cite{DBLP:conf/ijcai/KowalskiS85} classified legal rules, applying logic programming to model British laws. Golshani \cite{DBLP:journals/ijis/Golshani91} emphasized argument construction in automated legal reasoning while a more recent work by Sartor et al. \cite{DBLP:conf/iclp/SartorDBCPK22} utilized Logical English and top-down ASP solvers for legal encoding.

In image recognition, the YOLO network \cite{redmon2016you} excels in real-time object detection, such as applied in autonomous driving \cite{huang2022rd}. Of particular interest are instead, the study by Chen et al. \cite{chen2024multi} and the work by Yang et al. \cite{yang2024research} where YOLO is applyed for cell and cancer detection.

Summing up, the usage of ILP in weather prediction provides significant advantages over other methods. Specifically, ILP allows learning concepts that can be interpretable and adaptable to changing conditions. Moreover, unlike related works in the legal and image recognition domains, which primarily focus on statistical and deep learning approaches, our application of ILP introduces a level of transparency and interpretability that these other techniques lack. For example, in the legal domain, ILP can model complex rules and account for vagueness, while in image recognition, it complements neural networks by providing explainable reasons behind classification results, a feature rarely seen in standard approaches.

\section{Research goals}

The main goal of my research is to integrate deep learning with logic programming to create explainable AI systems. This involves leveraging logic programming, particularly Inductive Logic Programming (ILP), to model complex systems and achieve results that exhibit both high accuracy and explainability. When ILP is not able to reach high levels of accuracy, I want to employ deep learning system for the prediction tasks, due to its superior performance, followed by the application of logic programming to develop systems that can explain their decisions. 

In weather prediction, we are employing a hybrid learning approach, where sub-symbolic models (such as RNNs) will handle pattern recognition and time-series forecasting, while ILP will apply symbolic reasoning to provide post-hoc explanations for the RNN predictions. This approach falls under the \emph{'Post-hoc Explainability'} category of the taxonomy \cite{calegari2020integration}. The objective is to develop a system that can analyze meteorological data and generate logic rules that explain them and then use them to explain the outcome of the RNN model (which has an higher accuracy). In the future, we do not exclude the possibility to integrate a third system that will use Natural Language Processing to generate text-based explanations of weather forecasts, similar to those provided by professional meteorologists.

In the legal field, I am employing ILASP to model and reason about complex legal rules and cases. This project aims to develop systems that offer transparent and understandable logic behind automated decisions, thus enhancing trust and transparency in legal AI applications. The final model should assist legal professionals by basing its outcomes on numerous criminal records and providing explanations for the reasoning behind the final judgement. In this field, deep learning techniques would likely be used not to predict outcomes (since laws can be easily translated into logic rules) but to automatically encode articles of the legal code, given the vast amount of data.

For image recognition of bull's spermatozoa, my current focus is on using the YOLO (You Only Look Once) network for object detection. Future work will involve integrating ILP to improve the explainability of these systems, ensuring that the AI not only identifies objects accurately but also provides understandable reasons for its classifications.

\section{Current status of research}
My ongoing research and current work in each area of interest will be outlined in this section, along with citations of our publications that have shown promising results.

\begin{figure}[ht]
        \centering
\includegraphics[width=0.3\linewidth,trim=600 40 600 150, clip]{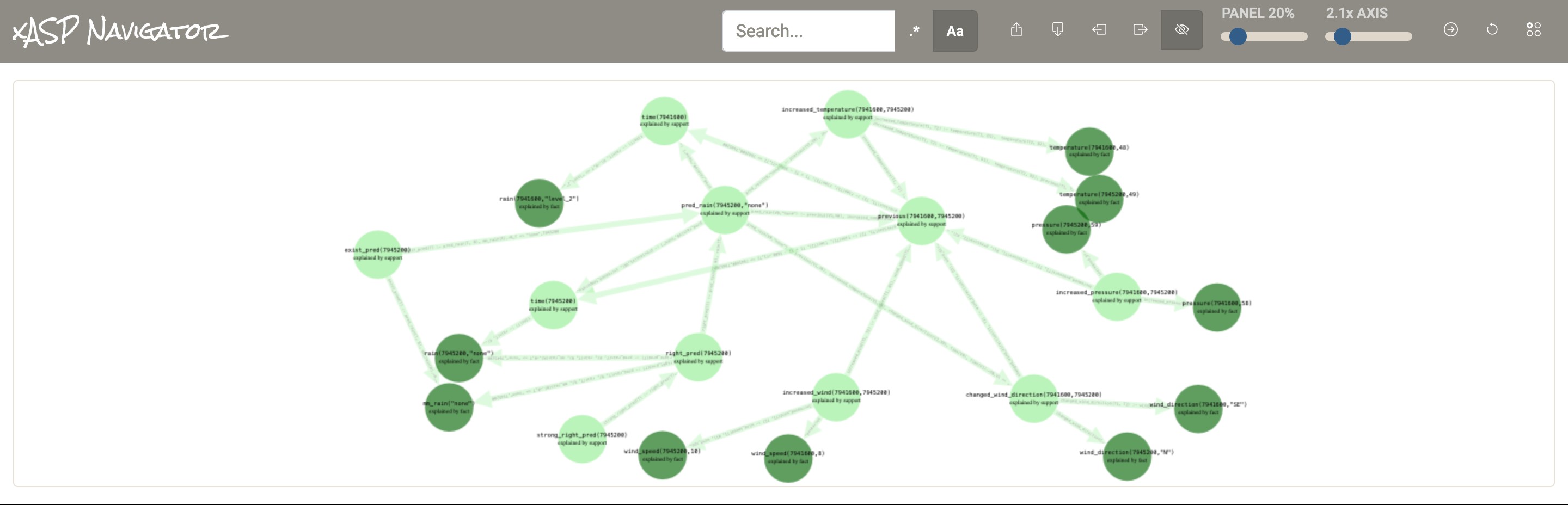} \hspace{1cm}
    \includegraphics[width=0.6\linewidth,trim=0 0 0 0]{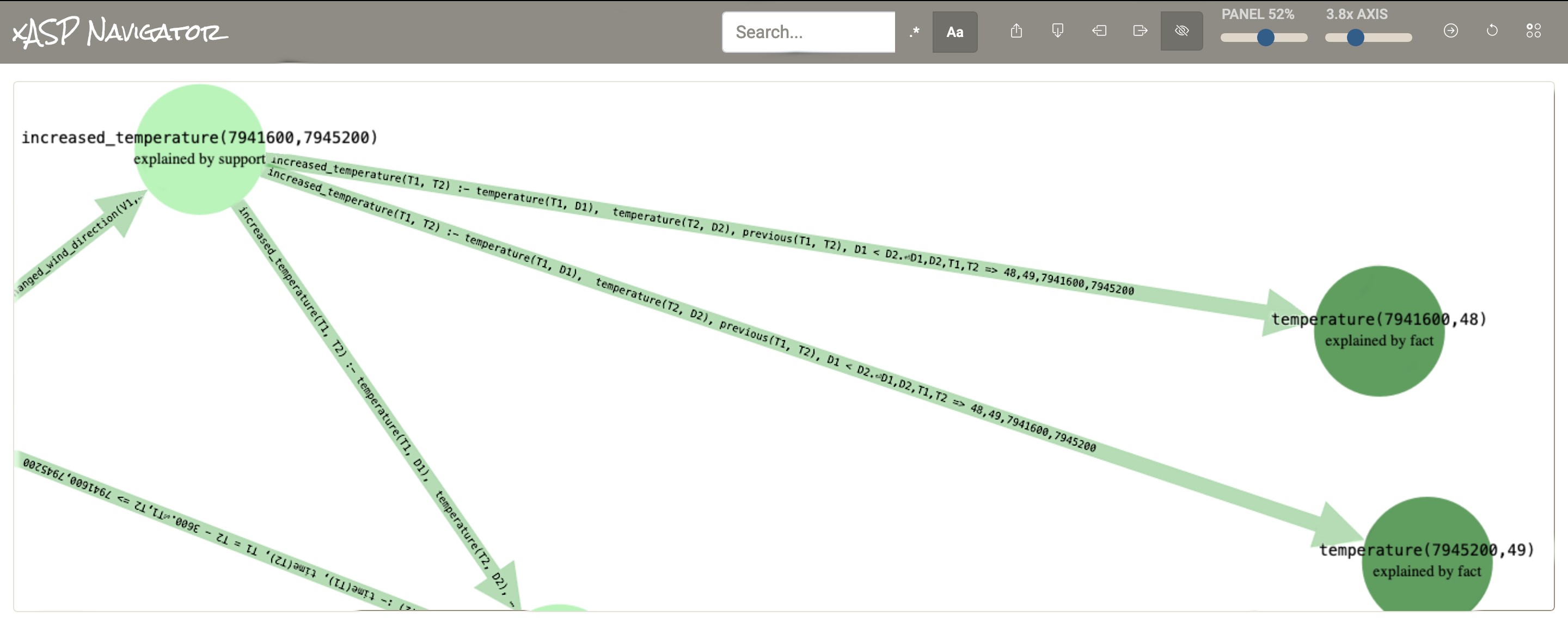}
    \caption{Full explanation of the answer set and zoomed-in view of a predicate explanation by xASP}
    \label{fig:zoom}
    \vspace{-7pt}
\end{figure}

In the domain of weather prediction, we have successfully developed a neural network capable of predicting the number of lightning strikes in the Friuli Venezia Giulia (IT) region with good accuracy. Initially, we used CNNs for this task and have now begun exploring recurrent neural networks to further enhance predictive performance. Our first results \cite{lpnmrpaper} indicate that the neural network performs well, though it requires some modifications to improve its robustness. Additionally, we have created a model using FastLAS that allows us to explain the predicted rainfall levels for the upcoming hours. Data was collected for two sets of three months each. Key atmospheric variables recorded include rain, temperature, humidity, wind speed, and pressure. The training process uses a 10-fold cross-validation approach, with each fold containing four days of training data and the remaining days used for validation. The hypothesis space is designed to include rules referencing past conditions to predict future states. This approach encourages the use of predicates indicating changes over time, while limiting the complexity of rules to maintain efficiency. In facts, we prioritizes rules that incorporate past information, using a scoring function to penalize predicates that do not reference multiple timestamps. The results of the experiments using FastLAS have been compared to other models such as SVM, RandomForest, and Decision Tree: although the accuracy of this system is not yet optimal, our initial findings indicate that FastLAS can be effectively used for this purpose since it can often reach the same accuracy as the other systems. The application of xASP is shown in Fig. \ref{fig:zoom}.

In the legal field, we have made significant progress by modeling four articles of the Italian Constitution and more or less a hundred precedent cases. This progress ensures that our system can adapt and improve its reasoning capabilities letting ILASP learn from historical legal cases. While the application of ILASP has not been wide so far, it provides a solid basis for further exploration for explainability \cite{d2020towards}. Additionally, we are addressing the inherent challenge of vagueness. This involves developing methods to handle ambiguous and context-dependent terms within legal texts, ensuring that our models can reason accurately despite these complexities. To address this issue we construct a model that enables the user to get all the different combination a vague concept can led to, exploiting choices rules. The evaluation of this model showed that it was successfully able to capture the legal distinctions and provide accurate classifications of cases. The results were validated against a set of real-world legal cases, demonstrating the model's ability to interpret complex legal scenarios and make decisions aligned with legal reasoning. Moreover, some incoherence and discrepancies in previous cases were discovered thanks to the system. The ASP model's output included detailed explanations for each classification, highlighting the rules and criteria applied. We presented our preliminary results in 
\cite{maast,dreossisemi, cilc2024}.

\begin{wrapfigure}{r}{0.35\textwidth}
    \centering
    \vspace{-10pt}
\includegraphics[width=0.35\textwidth,trim=20 50 45 50, clip]{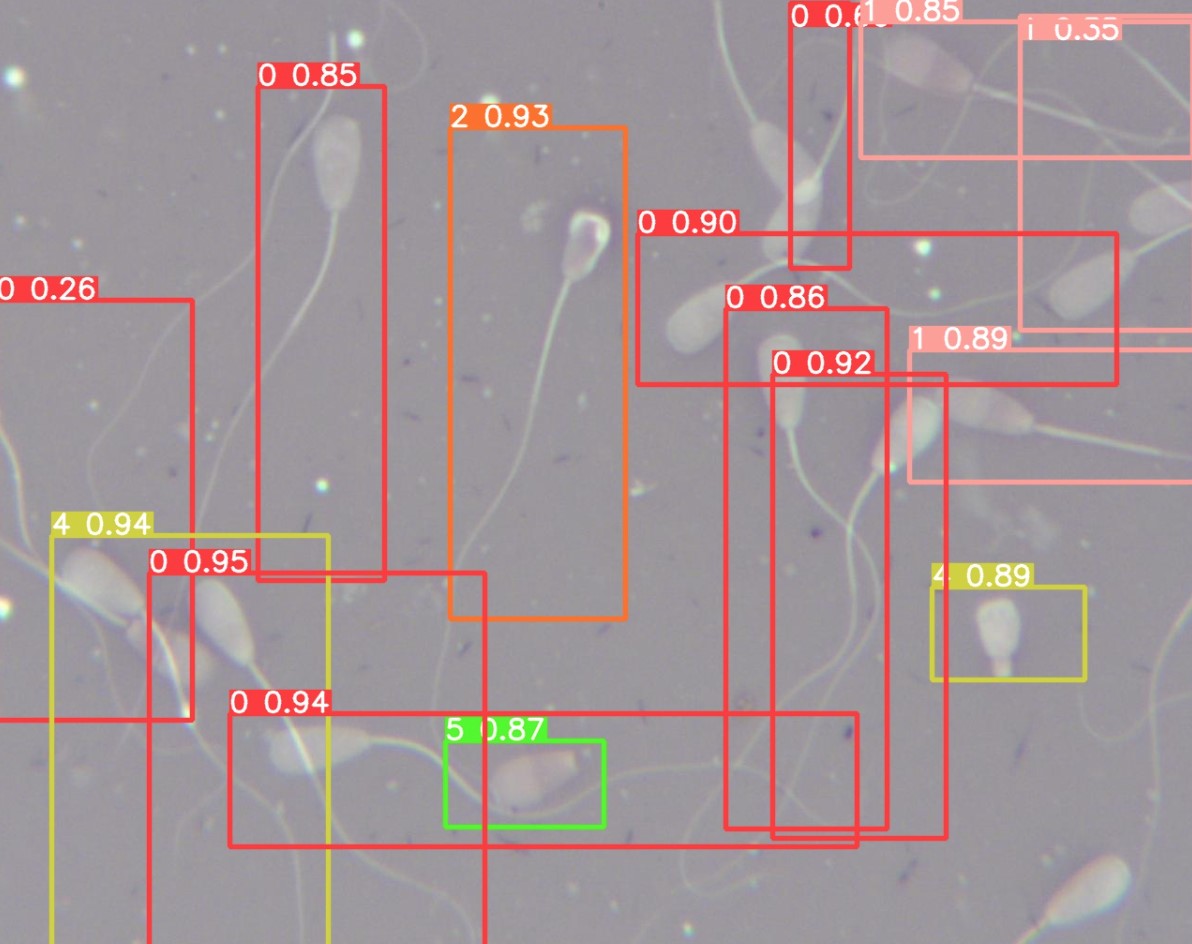}
    \caption{YOLO result on bull's spermatozoa image}
    \vspace{-30pt}
    \label{fig:spermatozoa}
\end{wrapfigure}

In both weather and legal domain, we are integrating explainability systems, specifically the one developed by Alviano et al. \cite{alviano2023advancements}, to ensure that our models provide transparent and understandable outputs. Indeed xASP is able to generate directed acyclic graphs which are particularly useful for representing dependencies and causal relationships within a logic program. This could be used in the future as a base to generate text that explain the outcomes. 

The image recognition in our research is still at the beginning stages, but you can see an example of detection our system is currently able to perform in Fig.\ref{fig:spermatozoa}. We have developed a basic neural network (using YOLO) that, despite its simplicity, is yielding good results (the accuracy of the current model reached 68\%). Our next step is to make these systems more reliable by integrating ILASP, as we already explored how it works \cite{dreossi2023exploring}. This integration aims to enhance the explainability and robustness of our image recognition systems.

 \vspace{-3mm}
\section{Conclusions}

This research project investigates the integration of deep learning and logic programming to develop AI systems that combine accuracy with explainability. By leveraging the strengths of neural networks for predictive tasks and ILP tools like FastLAS and ILASP for logical reasoning, the study aims to create interpretable models across various domains.

In weather prediction, the hybrid approach of combining FastLAS with recurrent neural networks aims to deliver accurate forecasts and clear, human-readable explanations, thus enhancing trust in the predictions. In the legal domain, ILASP is employed to model and reason about complex legal rules, providing transparent logic behind decisions and aiding legal professionals by offering understandable explanations for judgments. Finally, for image recognition, integrating YOLO networks with ILP improves AI interpretability, ensuring accurate object detection with clear reasoning.

Additionally, we are working towards making FastLAS compatible with GPU acceleration \cite{DBLP:journals/jetai/PaluDFP15,DBLP:conf/padl/DovierFPV16,DBLP:books/sp/18/DovierFP18,DBLP:journals/tplp/DovierFGHPR22} to enhance computational efficiency and scalability in AI applications. The pursuing of the research will hopefully bring advance in the field of AI and explainable AI.

\vspace{-0.4cm}

\setstretch{0.95}{
     {\footnotesize
    \paragraph{Acknowledgments.}
    This research is partially supported by Interdepartment Project on AI (Strategic Plan UniUD–2022-25), by NextGenerationEU-PNRR project \emph{MaPSART}-``Future Artificial Intelligence Research'', and by INdAM-GNCS 2024 project LCXAI: Logica Computazionale per  eXplainable Artificial Intelligence.}}
\vspace{-10pt}
\bibstyle{unsrt}

\bibliographystyle{eptcs}
\bibliography{generic}

\end{document}